Research Article

Bahadır Utku Kesgin, and Uğur Teğin[*]

# Photonic neural networks with spatiotemporal chaos in multimode fibers

**Abstract:** Optical computing has gained significant attention as a potential solution to the growing computational demands of machine learning, particularly for tasks requiring large-scale data processing and high energy efficiency. Optical systems offer promising alternatives to digital neural networks by exploiting light's parallelism. This study explores a photonic neural network design using spatiotemporal chaos within graded-index multimode fibers to improve machine learning performance. Through numerical simulations and experiments, we show that chaotic light propagation in multimode fibers enhances data classification accuracy across domains, including biomedical imaging, fashion, and satellite geospatial analysis. This chaotic optical approach enables high-dimensional transformations, amplifying data separability and differentiation for greater accuracy. Fine-tuning parameters such as pulse peak power optimizes the reservoir's chaotic properties, highlighting the need for careful calibration. These findings underscore the potential of chaos-based nonlinear photonic neural networks to advance optical computing in machine learning, paving the way for efficient, scalable architectures.

**Keywords:** Optical computing; Nonlinear optics; Optical fibers.

## 1 Introduction

In recent years, the explosive growth of machine learning applications has underscored the limitations of traditional, transistor-based computing systems, particularly as neural network models become more computationally demanding [1], [2], [3]. From image recognition to natural language processing, the colossal scale of data and parameters involved has driven a search for novel hardware solutions capable of handling these models' complexity without compromising efficiency [4], [5], [6]. Unlike digital systems that rely on electrical charge manipulation, optical systems leverage the unique properties of light and fundamental relations to intrinsically process information, making them particularly promising for applications in artificial intelligence and deep learning [7], [8], [9]. Thus, optical computing has emerged as a powerful candidate, offering intrinsic parallelism, energy efficiency, and the potential for high-speed data processing.

In recent years, advances in optical neural networks have highlighted the potential of light-based systems to perform fundamental neural network operations with programmable scattering media, free-space optics and integrated optical systems [10], [11], [12], [13], [14]. Recent studies of optical computing have demonstrated remarkable potential as photonic neural networks with scalable nonlinear manipulation of data through waveguides to efficiently implement complex nonlinear functions [15], and a programmable architecture has been demonstrated [16]. Furthermore, in recent years, nonlinear propagation-based optical computing in exotic waveguides and photonic crystal fibers has been proposed [17], [18]. These systems exploit nonlinear projection, wavefront and pulse shaping, and tailored nonlinear dynamics to achieve robust data transformations, achieving performance benchmarks that parallel or exceed traditional computational methods. Such advances underscore the potential for optical computing to meet and redefine the scalability of machine learning infrastructure. In addition to leveraging nonlinear dynamics for alternative computing schemes, chaos-based computing platforms demonstrated their potential with analog electronic [19], [20], [21], [22] and optical implementations [23], [24].

[*]Corresponding author: **Uğur Teğin**, Koç University Department of Electrical and Electronics Engineering, İstanbul, Türkiye; utegin@ku.edu.tr; https://orcid.org/0000-0002-4690-588X
**Bahadır Utku Kesgin:** Koç University Department of Electrical and Electronics Engineering, İstanbul, Türkiye; bkesgin22@ku.edu.tr; https://orcid.org/0009-0002-7262-3656

This work investigates a new pathway in optical fiber-based photonic neural network designs, leveraging spatiotemporal chaos in high-order modes of graded-index multimode fibers [25] to enhance the nonlinear optical processing. Through numerical analysis and experimental validation, we present a novel method for fiber-based optical neural networks to improve performance on the machine learning tasks. We demonstrate that spatiotemporal chaotic light propagation is advantageous for datasets that require extensive feature differentiation. We present significant improvements in the photonic neural network performance across various datasets, including the Breast MNIST, Fashion MNIST, and EuroSAT datasets. Our work underlines the potential of combining chaotic and nonlinear dynamics in optical waveguides for photonic neural networks to expand the capabilities of optical computing for machine learning, offering a novel alternative to digital neural networks.

## 2 Materials and Methods

### 2.1 Numerical studies

To investigate the spatiotemporal nonlinear pulse propagation in the multimode fibers we performed numerical studies and study spatiotemporal chaos-based optical computing architecture with time-dependent beam propagation method (TBPM) simulations. We implemented TBPM with graphical processing unit (GPU) in Python to simulate sufficiently fast nonlinear pulse propagation in the fiber.

$$\frac{\partial A}{\partial z} = \frac{i}{2k_0}\nabla^2 A - \frac{i\beta_2}{2}\frac{\partial^2 A}{\partial T^2} + \frac{\beta_3}{6}\frac{\partial^3 A}{\partial T^3}$$

$$-\frac{ik_0\Delta(x^2+y^2)}{R^2}A + i\gamma|A|^2 A$$

Here $A(x, y, z, T)$ is the slowly varying envelope of a multimode laser field, $z$ is the propagation axis, $T$ is the retarded time, $k_0 = \omega_0 n_0/c$ is the wave number where $\omega_0$ is the center frequency and $n_0$ is the refractive index of the center of the waveguide. Simulations using the time-dependent beam propagation method often demand extensive computational time due to the intensive multidimensional fast Fourier transform calculations involved. Our approach utilizes a careful scaling of both the propagation distance and pulse power to minimize computation time, while ensuring the essential nonlinear characteristics were preserved.

The launched pulses centered at 1030 nm with 1 ps duration were propagated for 5.5 mm equals to 10 self-imaging periods inside the graded-index multimode fiber. To create similar nonlinear strength with the experiments we set the peak power of the pulses to 1-20 MW range. This is a commonly used numerical trick in the literature to reach manageable computing time for spatiotemporal nonlinear pulse propagation simulations [15]. The time windows of the simulations are 20 ps with 9.8 fs resolution and the spatial windows is set to 54 by 54 um with 64x64 spatial grid.

### 2.2 Experimental studies

We use a picosecond Q-switched microchip laser (Thorlabs QSL103A) that produces 500 ps pulses with a repetition rate of 9 kHz for the light source in the experiments. The spectrum of the output pulse from the light source is centered at 1030 nm with a width of 1nm. Utilizing waveplates and a polarizing beamsplitter, we ensure that the output Gaussian beam is linearly polarized and the launched power to the fiber is controllable. Phase modulation to the Gaussian beam is performed using a phase-only spatial light modulator (SLM) (Holoeye Pluto-2.1 NIR-145), which has a pixel pitch of 8 μm and a 60 Hz refresh rate. Gaussian beam illuminates the 1080 pixels to 1080 pixels central region of the SLM. A blazed grating is added to the SLM phase to eliminate the unmodulated light from the fiber's path. An orbital angular momentum (OAM) phase with a charge of 1 is employed to create vortex beams that will excite high-order modes.

By superposing the grating, OAM, and information phases and converting them to 8-bit values with modulo 2π for projection, we complete our beam shaping process and encode the samples of the test datasets on the high-power laser beam. While all SLM surfaces are used for the vortex and grating phase, we only use the 800x800 region of the modulator with information to ensure that complete information is coupled into the fiber. All datasets used in the experiments are upsampled from their original dimensionality to 800x800. Five meters of a commercially available graded-index multimode fiber with 62.5 $\mu$m core size and a numerical aperture (NA) of 0.275 is used for the experiments. According to these parameters, our fiber supports 344 modes for single polarization. After 30 cm of free-space propagation, phase-modulated light is imaged onto the fiber in the focal point of a lens with a 35mm focal length. The fiber output is collimated using a lens with a focal length of 8mm. After collimation, output beam profiles are recorded using a 7.84 $\mu$m pixel pitch camera. The camera settings are kept constant throughout the experiments.

After optical information processing in the fiber, dimensionality reduction by average pooling is applied to 300x300 images, and 60x60 images are created for the samples of the test datasets. These images are then flattened to 1x3600 arrays to be used in the readout layer, which is a simple Ridge classifier to complete the machine learning task. In the ridge classifier, features are mapped to the [-1,1] range using L2 regularization, which prevents overfitting. After regularization, a simple linear layer completes classification. Default parameters of the Python's Scikit-learn package are used for the readout layer. A ratio of 80% training test and 20% test set are used to divide the datasets. A random state of 42 is used for all experiments for repeatability purposes. Confusion matrices are normalized row-wise (over accurate predictions), and decimals of normalized numbers are rounded.

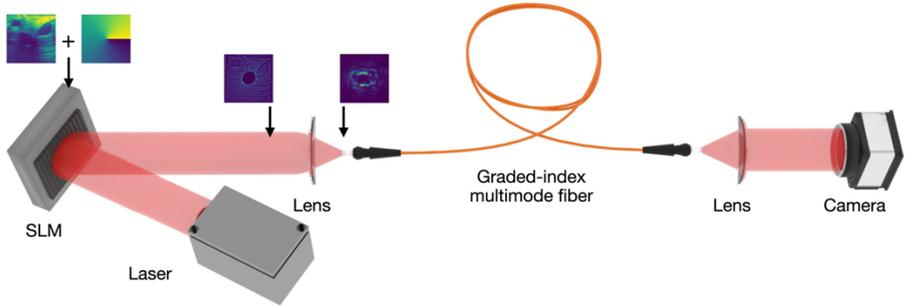

**Fig. 1:** Experimental setup. SLM, spatial light modulator. Input phase patterns and the beam profiles after the free-space propagation and focusing are indicated.

## 3 Results

In this study, a graded-index multimode fiber is utilized to allow a high spatial degree of freedom and complex nonlinear propagation without significant modal dispersion. Although these properties are useful for reservoir computing, when parameters are not optimized properly, nonlinear dynamics such as beam cleaning can highly decrease the overall efficiency of the model [15]. Specifically for this study, besides preventing nonlinear beam cleaning, we have to trigger spatiotemporal chaos with energy flow between the modes. As chaotic systems are sensitive to initial conditions, it is important to optimize launch parameters further, such as peak power. When these parameters are not optimized, data can be over- or under-processed within the multimode fiber, such that maximum performance cannot be achieved. Both in numerical and experimental studies, we optimize our system parameters for binary classification tasks before moving on to more complicated tasks.

### 3.1 Numerical studies

First, we employ numerical studies to assess the effectiveness of our chaotic reservoir compared to known models with GRIN MMF and to determine the effect of peak power on the system. We use the Breast MNIST dataset, composed of breast ultrasound images, to characterize the type of imaged tumor, benign or malignant. For the simulations of the stable state, we modulated the phase of an input Gaussian beam only with the phase of the information as done previously. To simulate the chaotic state, we modulated the phase of an input Gaussian beam with the phase of the information and the orbital angular momentum (OAM) phase. Phase modulation by the OAM phase leads to a vortex beam shape, where intensity accumulates in the ring rather than accumulating in the center. As initial energy is around the ring, we excite high-order modes. Due to the initial energy of high-order modes being significantly high compared to fundamental modes being significantly high, we satisfy conditions for chaos in the modal energy flow attractor. In both simulations, we couple light into a multimode fiber after phase modulation and propagation through the lens. We then record output beam profiles and apply ridge classification to complete learning.

Simulations with 1 MW, 5 MW, 10 MW, and 20MW of input peak powers are performed to assess system response in stable (without OAM phase) and chaotic regimes (with OAM phase). Here, we would like to emphasize that these peak powers are scaled to MW range to simulate less propagation distances thus to reduce simulation time. In our simulations we observe that the chaotic regime results in unorthodox outcomes compared to a stable regime. In 1 MW, due to low nonlinearity, neither encoding method is significant compared to the other. In 5 MW, the chaotic regime outputs the best accuracy of 83.34%, while the stable model can achieve a maximum accuracy of 82.70% at 10 MW. Results of the numerical simulations demonstrate that higher accuracies are feasible with less peak power compared to a stable state. After the optimal point, the accuracy of the chaotic model fluctuates due to over-processing. Similar effects on chaotic reservoirs have been demonstrated in the literature [22]. Simulation results and confusion matrices for the spatiotemporal chaos-based photonic neural network are illustrated in Figure 2.

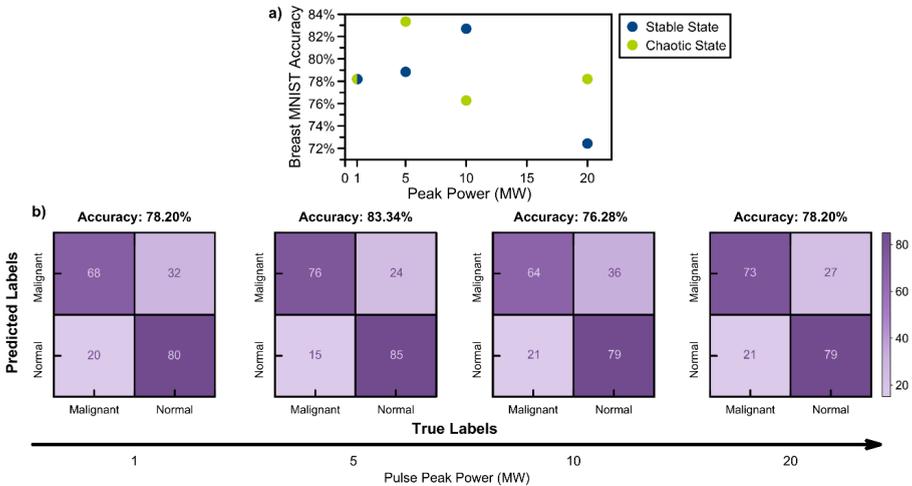

**Fig. 2:** Numerical simulation results on Breast MNIST dataset. **a.**, Recorded ridge classification accuracies of stable and chaotic regimes on different pulse peak powers. **b.**, Confusion matrices of ridge classification after chaotic transformation in different peak powers.

### 3.2 Experimental studies

#### 3.2.1 Experimental framework and selection of parameters

Similar to our numerical studies, we start the experiments by optimizing our pulse peak power for the reasons mentioned earlier that affect information processing and model performance. Again, we employ the Breast MNIST dataset to establish initial benchmarks in peak power. In the remainder of this study, we use the following encoding procedure. As the datasets we employ have small dimensionality, we up-sample all images to 800x800 and encode information to the input beam as a phase along with an orbital angular momentum phase using a spatial light modulator (SLM). After sending information to high-order modes and chaotic mode coupling within the multimode fiber, we obtain a speckle pattern on which we then employ ridge classification. After creating the pipeline of data encoding, we tested our model with various peak power settings.

With an input pulse peak power of 1 kW, our model performs at 78.84%. Although the model surpasses the baseline accuracy of 75%, data is under processed, and increasing peak power and nonlinearity is necessary. After increasing the pulse peak power to 2 kW, the model converges into its maximum recorded accuracy of 83.34%. Finally, when we increase the pulse peak power to 2.5 kW, although beam cleaning is not visible like in numerical studies, data is over-processed, which harms overall learning accuracy. The results of the peak power test are illustrated in Figure 3.

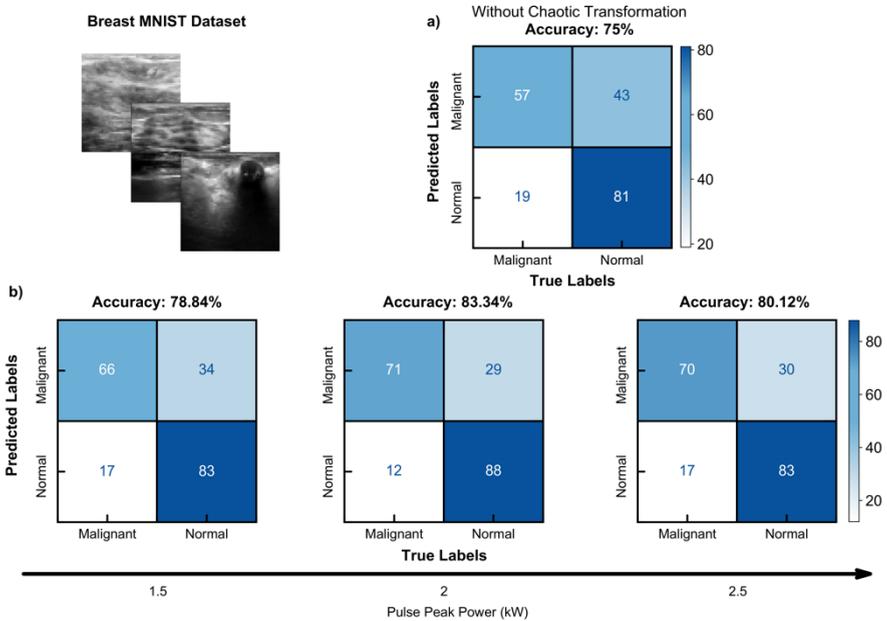

**Fig. 3:** Binary classification results utilizing Breast MNIST dataset. **a.** Confusion matrix of baseline ridge classification accuracy without chaotic transformation. **b.** Confusion matrices of ridge classification after chaotic transformation with different input pulse peak powers.

### 3.2.2 Breast MNIST Dataset

To assess the initial performance of machine learning with spatiotemporally chaotic photonic reservoir computing, we employ a binary classification before moving on to complex multi-class datasets. Breast MNIST dataset, which consists of breast ultrasound images taken from patients without a tumor or benign tumor or with a malignant tumor. We employ the same methodology in data processing and process our data again with a ridge classifier. Without our chaotic reservoir, the linear ridge classification layer results in a test set accuracy of 75%. As shown in Figure 3, we obtain a test set accuracy of 83.34%, improving baseline by 8% overall accuracy and by 14% malignant tumor accuracy. Our model performs on par with contemporary convolutional neural networks while training approximately seven orders of magnitude fewer parameters, as shown in Table 1. Such performance demonstrates the effectiveness of chaotic reservoir computing utilizing multimode fibers in biomedical disease characterization settings.

### 3.2.3 Fashion MNIST Dataset

Encouraged by the results in biomedical binary classification, we employ multi-class classification tasks with our computing method. The fashion MNIST dataset comprises 10 classes of various clothing types, such as T-shirts, bags, or boots. We use the official dataset split of 60,000 training and 10,000 test sets. We employ the same data encoding process to the SLM, record the corresponding output speckle patterns, and complete the data classification using linear readout. In this dataset, without chaotic transformation, the Ridge classification layer results with a complete test set accuracy of 75.82%. After the chaotic reservoir, the same readout algorithm results with a complete test set accuracy of 78.62%. As shown in the confusion matrices in Figure 4, we observe an increase in accuracy in most classes. Our framework performs better for this dataset than similar models implemented in multimode fibers, demonstrating the efficacy of chaotic elements in machine learning.

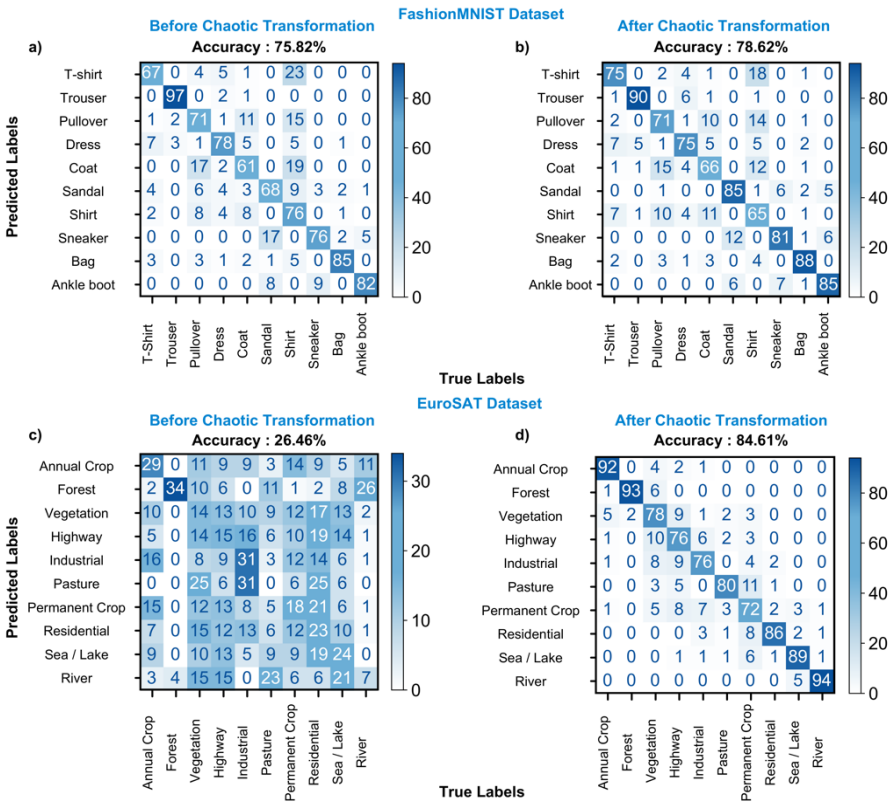

**Fig. 4:** Confusion matrices of ridge classification before and after chaotic transformation. **a,c.** Confusion matrices before chaotic transformation for FashionMNIST and EuroSAT dataset. **b,d.** Confusion matrices after chaotic transformation for FashionMNIST (b) and EuroSAT (d) dataset.

### 3.2.4 EuroSAT Dataset

To demonstrate a wide range of implementation for our model, we decided to move on to a geospatial dataset that utilizes satellite images. For this task, we use the EuroSAT dataset, composed of 10 classes of satellite imagery, including industrial areas, residential areas, and different crop areas. After following the same data processing and collection procedure, we again employ ridge classification. For the EuroSAT dataset, baseline ridge classification without our chaotic reservoir outputs an accuracy of 26.46%, demonstrating the task's difficulty and linear inseparability of samples in feature space. After processing input data using our chaotic reservoir, we observe a 58.15% increase in accuracy from 26.46% to 84.61. Confusion matrices in Figure 4 also demonstrate the transition from scrambled data to linearly separable data, along with the significant increase in accuracy in all classes. Similar to the BreastMNIST dataset, we compare the performance of our model with modern convolutional neural networks.

Our method works on par with convolutional neural networks (CNN), again training many orders of magnitude fewer parameters. Our benchmark CNN models are not fine-tuned over many hours of training but are trained from scratch. These results demonstrate the efficiency of the spatiotemporal chaotic information processing method on an experimental basis.

| Model | Number of Trainable Parameters | BreastMNIST Accuracy | EuroSAT Accuracy |
|---|---|---|---|
| Proposed | 3600 | 83.34% | 84.61% |
| ResNet-18 | 11,000,000 | 83.3% | 80.08% [26] |
| ResNet-50 | 25,000,000 | 84.2% | 87.33% [27] |

**Table 1:** Classification results of three different models on Breast MNIST and EuroSAT datasets.

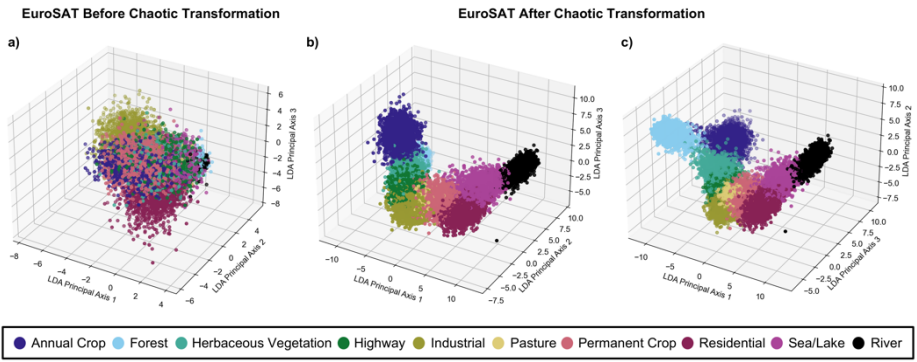

**Fig. 5:** Impact of chaotic nonlinear transformation on EuroSAT data points to enhance classification performance. **a.** Feature space before chaotic transformation. **b,c.** Feature space after chaotic transformation.

### 3.2.5 Clustering property of spatiotemporal chaotic reservoir

Previous works in the literature demonstrate pattern recognition and clustering by utilizing chaotic dynamical systems as reservoirs [28]. In this study, we further analyze the EuroSAT dataset's results and observe our chaotic reservoir's clustering properties. We perform linear discriminant analysis (LDA) on raw and processed data to visualize three-dimensional feature space. In the initial distribution illustrated in Figure 5, it can be seen that the classes cannot be separated with any linear function, demonstrating the complexity of the dataset. After chaotic transformation, data points cluster alongside other data points that belong to the same class, which results in easily separable data and significantly higher accuracies in all classes. While shedding light on classification results in Figure 4, these results also demonstrate the common effect of clustering among chaotic reservoir computers in our system.

## 4 Discussion

In this study, we explore the integration of spatiotemporal chaotic dynamics within graded-index multimode fibers (GRIN MMFs) to enhance the performance of fiber-based photonic neural network architectures. By employing both numerical simulations and experimental studies, we demonstrate that chaotic light propagation significantly improves classification accuracy across various datasets, including the Breast MNIST, Fashion MNIST, and Eurosat. The chaotic regime achieves enhanced dimensionality and separability, effectively leveraging high-order mode excitation and nonlinearity to increase model performance. This spatiotemporal chaos approach proved particularly advantageous in datasets that require extensive feature differentiation. Furthermore, we observe that optimizing parameters such as pulse peak power and initial energy distribution are crucial to maintaining high performance for machine learning tasks with spatiotemporal chaos-based photonic neural networks. Our proposed photonic computing architecture consumes less than 75 W power which is order magnitude less than a typical commercially available GPU for machine learning studies.

In conclusion, the use of spatiotemporal chaos within photonic neural networks provides a promising pathway toward energy-efficient, high-performance optical computing for machine learning tasks. Our study reveals that this architecture not only matches but often exceeds the capabilities of traditional digital and previously reported multimode fiber-based optical computing systems, while employing far fewer parameters. Additionally, this study highlights the clustering ability of chaotic optical computing systems, which significantly improves the linear separability of data in feature space. These findings open new avenues for developing sustainable and scalable optical computing hardware. Future work will focus on expanding the applicability of chaotic photonic reservoirs to more complex, real-world datasets and exploring alternative chaotic dynamics for broader computational tasks.

**Research funding:** This work is supported by the Scientific and Technological Research Council of Turkey (TÜBİTAK) under grant number 123F171.

**Author contribution:** All authors have accepted responsibility for the entire content of this manuscript and consented to its submission to the journal, reviewed all the results and approved the final version of the manuscript. B.U.K. performed simulations and tests, U.T supervised and directed the project. All the authors participated in the data analysis.

**Conflict of interest**: Authors state no conflict of interest.

**Data availability statement**: Data and code related to the results in this work may be obtained from the authors upon reasonable request.

**Supplementary Material**

Bahadır Utku Kesgin, and Uğur Teğin[*]

# Photonic neural networks with spatiotemporal chaos in multimode fibers

**Supplementary Notes 1: Stability of experiment and coupled power with vortex beam**

In nonlinear optical computing, the most important variable is the optical power within the system as nonlinear interactions depend on intensity. Fluctuations in the output power of the laser are expected due to non-ideal environmental factors and inevitable noise. To test the overall power fluctuations, we measured the power coupled to the fiber with a power meter before the experiments. Although the duration of the longest experiment was 6.4 hours (FashionMNIST [1]) we measured the fiber-coupled output power for 12 hours. Normalizing the output power by the maximum value of the measurements, the minimum coupled power was 0.9817. The duration of other experiments is 2.5 hours for EuroSAT [2] and 5 minutes for BreastMNIST [3].

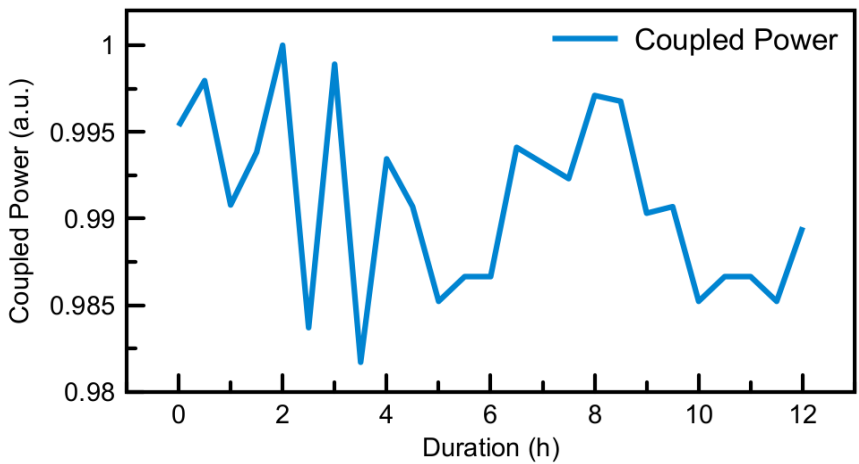

**Supplementary Figure 1:** Fiber coupled laser power.

**Supplementary Notes 2: Preliminary tests on spatiotemporal chaotic state with vortex beam**

In this study, we demonstrate an optical computing system based on the excitation of high order modes. When the initial energy of high order modes compared to fundamental modes is higher than 1/50, spatiotemporal chaos occurs, and fiber cannot converge into a stable state[4]. In high powers, the stable state refers to beam cleaning. We tested the impact of the vortex beam compared to the Gaussian beam with and without information encoding in 2kW peak power. As described in the results section, we added a blazed grating phase to prevent unmodulated zero order from entering the fiber. Due to the orbital angular momentum phase spot size of the bottle beam will be slightly larger than the regular Gaussian beam. We measured the power of fiber-coupled light with and without a vortex beam to test full coupling. Without the vortex beam, we measured an average power of 18.4mW and with the vortex beam, we measured an average power of 18.1mW. Since we encode information in an inner square, this coupling difference is negligible. If we increase the charge of orbital angular momentum phase, the spot size will also increase. Due to this we did not increase the charge further since it will lead to loss of coupled power.

[*]**Corresponding author: Uğur Teğin**, Koç University, İstanbul, Türkiye; utegin@ku.edu.tr; https://orcid.org/0000-0002-4690-588X
**Bahadır Utku Kesgin:** Koç University, İstanbul, Türkiye; bkesgin22@ku.edu.tr; https://orcid.org/0009-0002-7262-3656

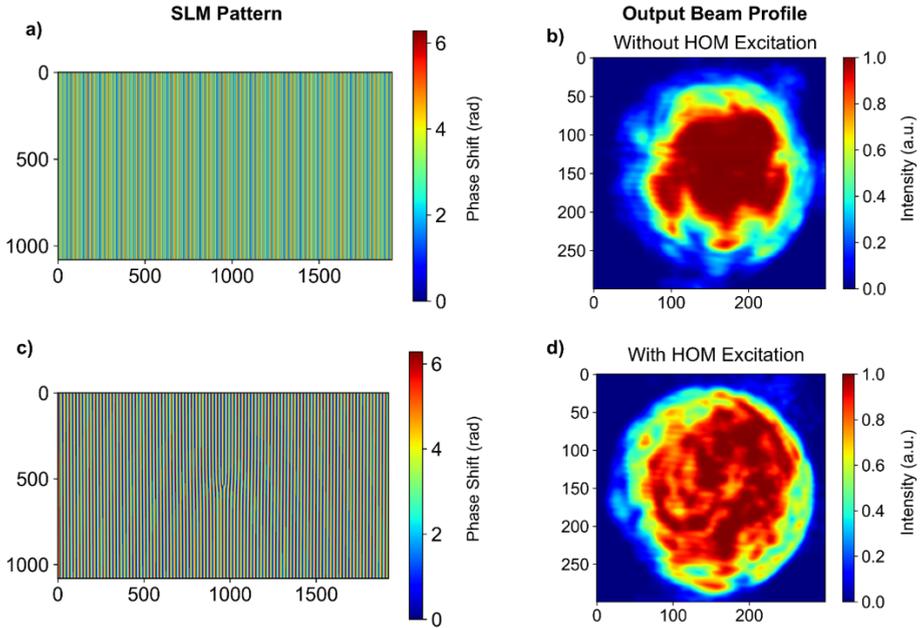

**Supplementary Figure 2:** Impact of vortex beams on the stability of the propagation. a. Regular modulation phase is applied to the input Gaussian beam. b. Output beam profile without high order mode excitation. c. The orbital angular momentum modulation phase was applied to the input Gaussian beam. d. Output beam profile with high order mode excitation.

Supplementary figures 2 and 3 demonstrate that without the excitation of high order modes system starts to converge to its stable state. By increasing the initial energy coupled to high order modes system cannot converge to its stable state verifying the numerical results and the presence of spatiotemporal chaos in multimodal propagation. As a final preliminary test, we sent different information patterns with vortex beams and verified our observations.

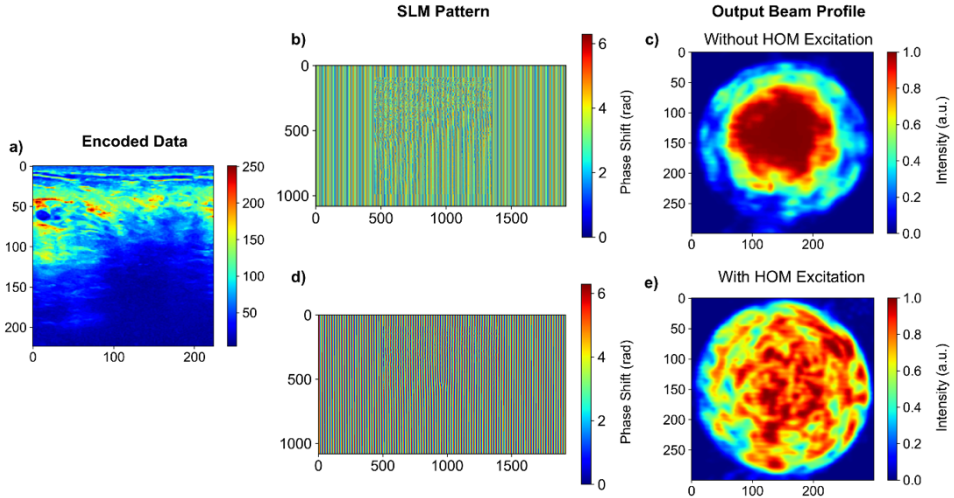

**Suppplementary Figure 3:** Impact of vortex beams on the stability of the propagation with data. a. Encoded data in phase patterns. b. Regular modulation phase is applied to the input Gaussian beam. c. Output beam profile without high order mode excitation. d. Orbital angular momentum modulation phase was applied to the input Gaussian beam. e.Output beam profile with high order mode excitation.

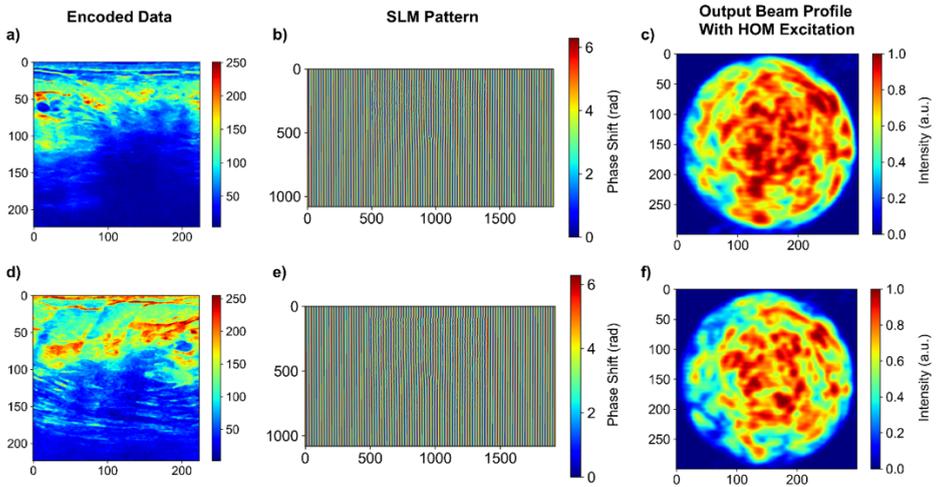

**Supplementary Figure 4:** Impact of vortex beams on the stability of the propagation with data. a,d. Encoded data in phase patterns. b, e. The orbital angular momentum modulation phase was applied to the input Gaussian beam. c,e. Output beam profile with high-order mode excitation.